\documentclass[prl,twocolumn,amsmath,amssymb]{revtex4-1}

\usepackage{graphicx}
\usepackage{bm}
\usepackage{gensymb}
\usepackage{multirow}
\usepackage{xcolor}
\usepackage{hyperref} 
\newcommand{\beginsupplement}{%
        \setcounter{table}{0}
        \renewcommand{\thetable}{S\arabic{table}}%
        \setcounter{figure}{0}
        \renewcommand{\thefigure}{S\arabic{figure}}%
     }

\begin{document}

\title{Ferromagnetic ordering along the hard axis in the Kondo lattice YbIr$_{3}$Ge$_{7}$}

\author{Binod K. Rai$^{1}$}
\email[]{These authors contributed equally to this work.}
\author{Macy Stavinoha$^{2}$}
\email[]{These authors contributed equally to this work.}
\author{J. Banda$^3$}
\author{D. Hafner$^3$}
\author{Katherine A. Benavides$^4$}
\author{D.~A.~Sokolov$^3$}
\author{Julia Chan$^4$}
\author{M. Brando$^3$}
\author{C.-L. Huang$^2$}
\author{E. Morosan$^{1,2}$}

\affiliation{$^1$ Department of Physics and Astronomy, Rice University, Houston, TX 77005 USA
\\$^2$ Department of Chemistry, Rice University, Houston, TX 77005 USA
\\\\$^3$Max Planck Institute for Chemical Physics of Solids, N\"{o}thnitzer Strasse 40,  Dresden, 01187, Germany
\\\\\\$^4$ Department of Chemistry, University of Texas at Dallas, Richardson, TX 75080, USA}
\date{\today}

\begin{abstract}
Ferromagnetic Kondo lattice compounds are far less common than their antiferromagnetic analogs. In this work, we report the discovery of a new ferromagnetic Kondo lattice compound, YbIr$_{3}$Ge$_{7}$. Like almost all ferromagnetic Kondo lattice systems, YbIr$_{3}$Ge$_{7}$ shows magnetic order with moments aligned orthogonal to the crystal electric field (CEF) easy axis. YbIr$_{3}$Ge$_{7}$ is unique in that it is the only member of this class of compounds that crystallizes in a rhombohedral structure with a trigonal point symmetry of the magnetic site, and it lacks broken inversion symmetry at the local moment site. AC magnetic susceptibility, magnetization, and specific heat measurements show that YbIr$_{3}$Ge$_{7}$ has a Kondo temperature $T_{\rm K}$~$\approx$~14~K and a Curie temperature $T_{\rm C}$ = 2.4 K. Ferromagnetic order occurs along the crystallographic [100] hard CEF axis despite the large CEF anisotropy of the ground state Kramers doublet with a saturation moment along [001] almost four times larger than the one along [100]. This implies that a mechanism which considers the anisotropy in the exchange interaction to explain the hard axis ordering is unlikely. On the other hand, the broad second-order phase transition at $T_{\rm C}$ favors a fluctuation-induced mechanism. 
\end{abstract}

\maketitle

Various competing ground states in Kondo lattice (KL) systems, governed by the delicate balance between the Ruderman-Kittel-Kasuya-Yosida (RKKY) exchange interaction and on-site Kondo effect, have gained great interest for over three decades \cite{Doniach,Stewart2001,Lohneysen2007}. These two interactions usually result in antiferromagnetic (AFM) order with a dense KL metallic ground state. The balance between Kondo and RKKY interactions can be tuned by applying non-thermal parameters such as pressure, magnetic field, or chemical doping, resulting in non-Fermi liquid behavior near a quantum critical point (QCP) where the AFM transition temperature is suppressed to absolute zero, or other quantum collective phenomena emerge including unconventional superconductivity \cite{Gegenwart2007,Paglione2007,Keimer2017}. 

Among known KLs, the number of compounds that shows AFM order greatly surpasses that of the ferromagnetically ordered compounds \cite{Brando2016,Ahamed2018,Hafner2018}. Thus, in stark contrast to the AFM counterpart, in-depth theoretical work to describe ferromagnetic (FM) KL compounds is largely missing \cite{Yamamoto2010,Kruger2014}. Recently Ahamed et al. \cite{Ahamed2018} suggested that broken inversion symmetry at the local moment site could promote FM order in KLs. While this scenario is possible for most of the Ce- and Yb-based FM KL compounds,  including CeTiGe$_3$ \cite{Fritsch2015}, YbNiSn \cite{Bonville1992}, YbPtGe \cite{Katoh2009}, YbRhSb \cite{Muro2004}, YbPdSi \cite{Tsujii2016}, and the most heavily studied FM KLs CeAgSb$_2$ \cite{Myers1999,Andre2000,Sidorov2003,Araki2003,Takeuchi2003}, CeRuPO \cite{Krellner2008,Krellner2007,Krellner2008relevance} and YbNi$_4$P$_2$ \cite{Krellner2011}, it does not apply to systems \emph{with} inversion symmetry like Yb(Rh$_{0.73}$Co$_{0.27}$)$_2$Si$_{2}$ \cite{Klingner2011,Lausberg2013} and YbCu$_2$Si$_2$ \cite{Shimizu1987,Tateiwa2014}. Moreover, it has been theoretically and experimentally found that the FM phase is inherently unstable, either towards a first-order phase transition \cite{Belitz1999} or towards inhomogeneous magnetic phases \cite{Kotegawa2013}. Thus, the FM QCP either does not exist or is masked by other phases. Only in the case of YbNi$_4$P$_2$, FM order occurs via a second-order phase transition upon chemical substitution in YbNi$_4$(P$_{1-x}$As$_x$)$_2$ \cite{Steppke2013}. The presence of a FM QCP in this system has been attributed to its quasi-1D structure \cite{Krellner2011,Steppke2013}. Thus, in order to develop the theory surrounding FM KL systems in general, and experimentally realize new FM QCPs in particular, new FM KL compounds are called for.  

In compounds with strong crystal electric field (CEF) effects, the CEF-induced anisotropy determines the direction of easy and hard magnetization axes in the paramagnetic (PM) state.  Interestingly, in all of the FM KL compounds mentioned above, with the exception of CeTiGe$_{3}$ and YbCu$_2$Si$_2$, this CEF anisotropy competes with the RKKY interaction and results in magnetic ordering along the axis \textit{orthogonal} to the CEF easy axis \cite{Myers1999,Krellner2008,Krellner2011, Krellner2012,Lausberg2013,Steppke2013}. Even more astonishing is the fact that the FM hard axis ordering appears to be a general trait of FM KL systems, as of yet unexplained. \cite{Hafner2018}. 


In this paper, we report the discovery of a new FM KL compound YbIr$_{3}$Ge$_{7}$ with $T_{\rm C}$ = 2.4 K. In line with the above empirical observation, spontaneous magnetization occurs along the hard CEF axis. However, YbIr$_{3}$Ge$_{7}$ is unique among FM KLs because it is the only such compound crystallizing in a rhombohedral lattice, and it does not show broken inversion symmetry at the local moment site. Recently, we discovered a series of Ce- and Yb-based compounds in this 1-3-7 structural family, including YbRh$_3$Si$_7$ \cite{Rai2018YbRh3Si7}, CeIr$_3$Ge$_7$ \cite{Rai2018CeIr3Ge7,Banda2018}, and YbIr$_3$Si$_7$ \cite{StavinohaYbIr3Si7}. Among these compounds, YbRh$_3$Si$_7$ and YbIr$_3$Si$_7$ are KL compounds with the former proposed to order antiferromagnetically below 7.5 K based on neutron scattering, and the latter showing ferromagnetic correlations below 4 K, with the moments ordered along the hard CEF axis in both. In contrast, CeIr$_3$Ge$_7$ shows AFM order along the easy CEF axis at a remarkably low temperature $T_{\rm N}$ = 0.63 K, in the absence of Kondo screening or frustration. Although chemically and structurally similar, the balance between CEF effects, Kondo screening, and RKKY interactions in these systems differ drastically. Thus, this family of compounds presents an opportunity to study the delicate competition among these interactions and the resulting ground states.  


Single crystals of YbIr$_{3}$Ge$_{7}$ were grown using Ge self-flux, as described in earlier publications \cite{Remeika1980,Rai2015}. The purity and crystal structure were identified by single crystal and powder x-ray diffraction analysis (Tables \ref{SCXRD} and \ref{Atomic} and Fig. \ref{XRD} in the Appendix). YbIr$_{3}$Ge$_{7}$ crystallizes in the ScRh$_{3}$Si$_{7}$ structure type \cite{Chabot1981,Lorenz2006} with lattice parameters $a=7.8062(10)$\,\AA~and $c=20.621(5)$\,\AA. The stoichiometry determined by free variable refinement of the occupancies is YbIr$_3$Ge$_{7-\delta}$ where $\delta$ = 0.3. The crystals were oriented along the [100] and [001] hexagonal axes using a back-scattering Laue camera. Room temperature powder x-ray diffraction data were collected using a Bruker D8 diffractometer with Cu K$\alpha$ radiation, with additional room temperature single crystal x-ray diffraction performed using a Bruker D8 Quest Kappa single crystal x-ray diffractometer equipped with an I$\mu$S microfocus source, a HELIOS optics monochromator, and a CMOS detector. The anisotropic DC magnetic susceptibility $M/H$ and magnetization data were measured using a Quantum Design (QD) Magnetic Property Measurement System (MPMS) with an applied magnetic field up to 7 T. AC susceptibility was measured with a QD MPMS with a modulation field amplitude $\mu_0H_{\rm{ac}}=1$\,mT at a frequency of 113.7\,Hz. AC susceptibility measurements at 20 mK were performed using an Oxford Instruments dilution refrigerator. Specific heat and electrical transport measurements were performed in a QD Physical Property Measurement System.

\begin{figure}[b]
\includegraphics[clip,width=\columnwidth]{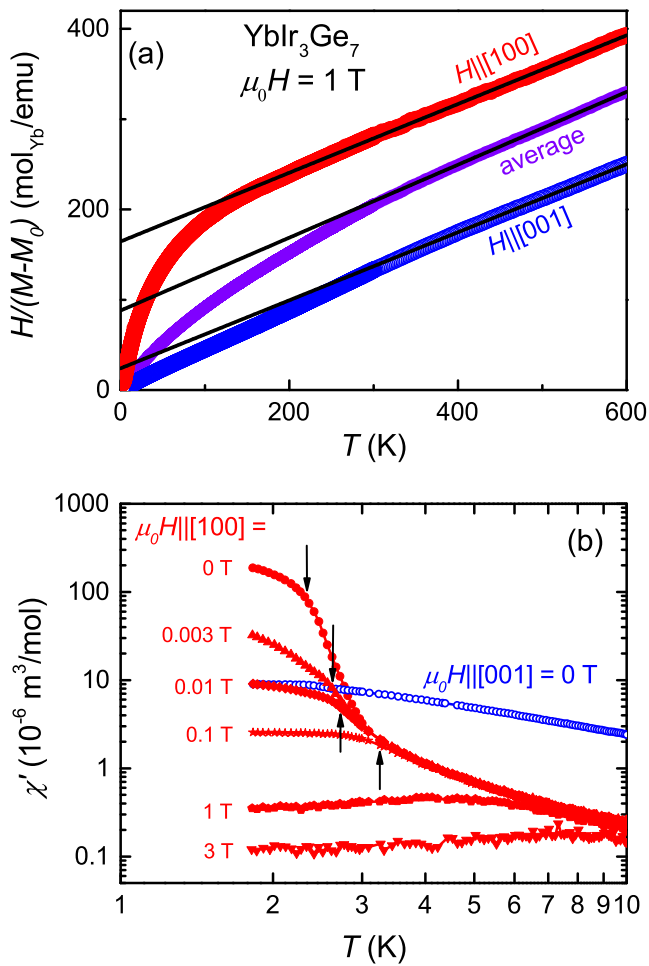}
\caption{\label{Susceptibility} (a) Inverse magnetic susceptibility $H/(M-M_0)$ vs. $T$ with the polycrystalline average, $M_{avg} = (M_{001} + 2M_{100})$/3 (purple line) and Curie-Weiss fits between $T$ = 400 - 600 K (solid black lines) (b) Temperature-dependent AC magnetic susceptibility $\chi'$ with $H_{\rm{ac}}$ along [001] (blue open symbols) and along [100] (red closed symbols) with different applied static fields $H$ applied also along [100].} 
\end{figure}

In YbIr$_{3}$Ge$_{7}$, the Yb atom occupies a trigonal point symmetry ($\bar{3}$), and the $J = 7/2$ energy levels are split by the CEF in four Kramers doublets. While the CEF energy levels for the Ce isostructural compound were determined from magnetic susceptibility measurements and calculations \cite{Rai2018CeIr3Ge7,Banda2018}, the larger angular momentum of the Yb leads to a corresponding Hamiltonian with six parameters in YbIr$_{3}$Ge$_{7}$, which can not be fully solved with the data at hand. However, large CEF anisotropy in YbIr$_{3}$Ge$_{7}$ is evidenced by the linear high-temperature inverse susceptibility $H/(M-M_0)$ shown in Fig. \ref{Susceptibility}a, measured for field $H \parallel [001]$ (blue symbols) and $H \parallel [100]$ (red symbols). The Curie-Weiss fit (solid line) of the average inverse susceptibility (purple) between 400 and 600\,K yields the experimental effective moment $\mu_{\mathrm{eff}} = 4.42\,\mu_{\mathrm{B}}$/Yb, close to the calculated value $4.54\,\mu_{\mathrm{B}}$/Yb for Yb$^{3+}$. The paramagnetic Weiss temperatures along [100] and [001] are both negative, $\theta_{\rm W}^{[100]} = -400$\,K and $\theta_{\rm W}^{[001]} = -70$\,K, and yield a first CEF parameter $B_{2}^{0} = -1.6$\,meV \cite{Bowden1971}, which is a measure of the strength of the CEF anisotropy. Deviations from linearity below 300 K indicate that CEF splitting exceeds this temperature range, similar to the large splitting observed in the Ce analog \cite{Rai2018CeIr3Ge7,Banda2018}.

Further insight into the low-temperature magnetic properties of YbIr$_{3}$Ge$_{7}$ comes from the magnetic AC susceptibility shown in Fig. \ref{Susceptibility}b. AC susceptibility measurements reveal spontaneous magnetization below $T_{\mathrm{C}} =$ 2.4~K for H$\parallel$ [100], indicative of ferromagnetic order. This is indeed consistent with the transition moving up in temperature (vertical arrows, Fig. \ref{Susceptibility}b) with increasing applied field. At zero field the susceptibilities measured with $H_\mathrm{ac}\parallel[100]$ and $H_\mathrm{ac}\parallel[001]$ cross each other at a temperature just above $T_\mathrm{C}$. This behavior is similar to that of the heavy fermion ferromagnet YbNi$_4$P$_2$ and of all other KL ferromagnets which order along the hard direction \cite{Steppke2013,Hafner2018}.

\begin{figure}[htbp]
	\includegraphics[clip,width=\columnwidth]{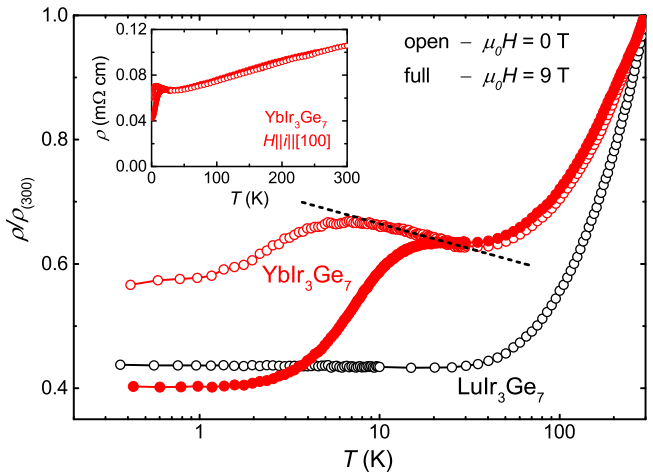}
\caption{\label{Res} Scaled temperature-dependent electrical resistivity $\rho/\rho_{300}$ of YbIr$_{3}$Ge$_{7}$ (red circles) with $\mu_0 H = 0$ (open symbos) and 9 T (closed symbols) for $H||i||$[100]. The nonmagnetic polycrystalline analog LuIr$_{3}$Ge$_{7}$ is shown with black symbols. The dashed line in the main panel shows a $-$ln$T$ increase in $\rho(T)$. The inset shows the absolute resistivity of YbIr$_{3}$Ge$_{7}$ with $\mu_0H$ = 0  and 9 T}
\end{figure}

\begin{figure}[h!]
\includegraphics[width=\columnwidth,clip]{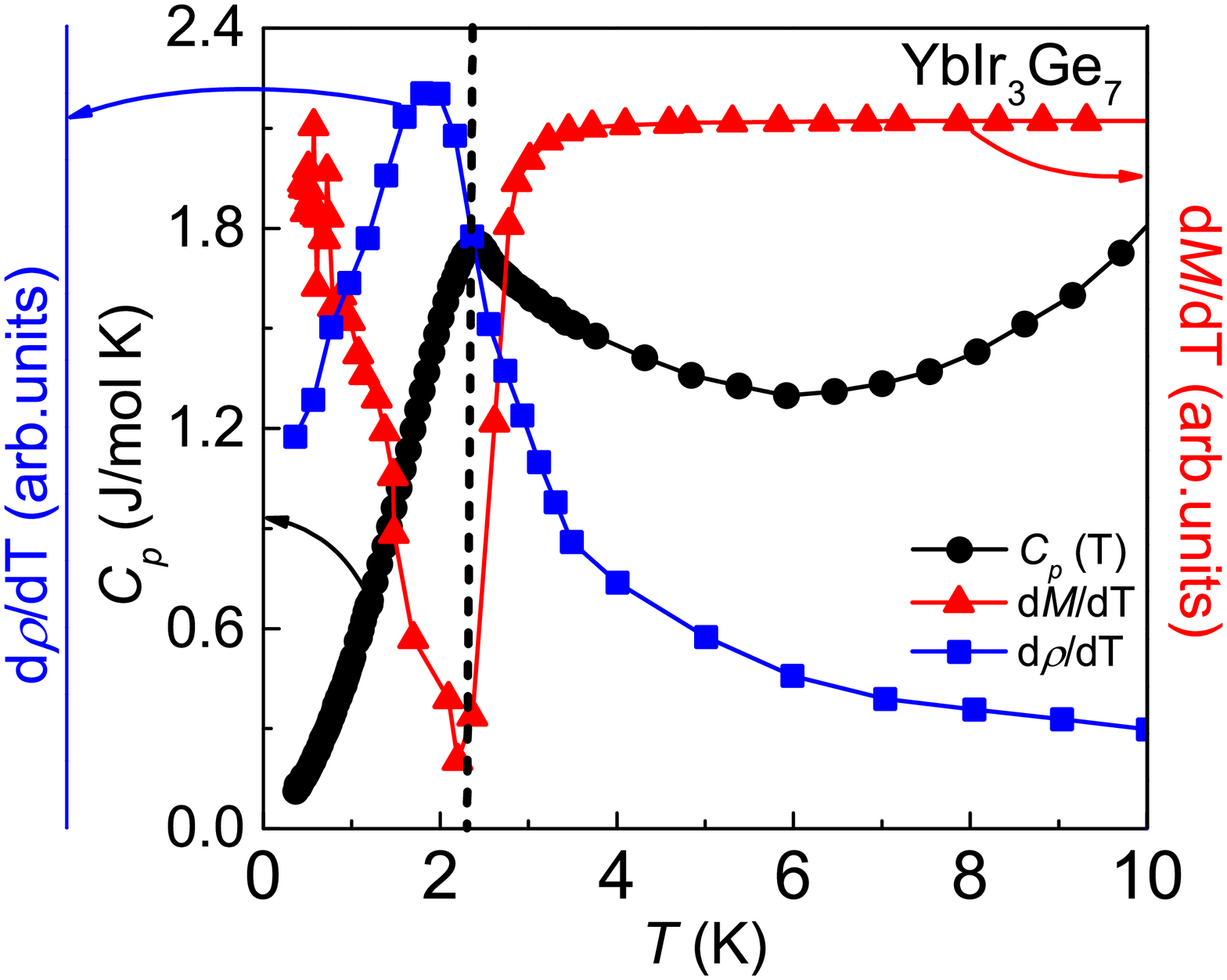}
\caption{\label{TC}The ordering temperature $T_{\rm C}$ (vertical dashed line) for YbIr$_3$Ge$_7$ determined from peaks in C$_p$ ($H$ = 0, black circles, left axis), a minimum in d$M$/dT ($\mu_0H$ = 0.1 T, red triangles, right axis), and d$\rho$/dT ($H$ = 0, blue squares, far left axis)}
\end{figure}

The temperature-dependent electrical resistivity in YbIr$_{3}$Ge$_{7}$ is typical of dense KL systems, as shown in Fig.~\ref{Res}. Metallic behavior with a positive resistivity coefficient (d$\rho$/d$T >$ 0) is observed between 300 K and 40 K. On further cooling, the resistivity shows a local minimum around 35 K, followed by a $-$ln$T$ increase (dashed line) down to a coherence maximum around 6 K for $H$ = 0 (red open symbols, Fig. \ref{Res}), reflecting the incoherent Kondo scattering behavior. A drop in resistivity is seen as the temperature is further lowered through a magnetic phase transition around 2.4 K, as shown more clearly in a derivative plot in Fig.~\ref{TC}. The resistivity in applied magnetic field $\mu_0H$ = 9 T (full symbols, Fig. \ref{Res}) shows the partial suppression of the Kondo effect and the FM fluctuations as the logarithmic increase disappears and the local maximum moves up in temperature. 

\begin{figure}[htbh]
  \includegraphics[clip,width=\columnwidth]{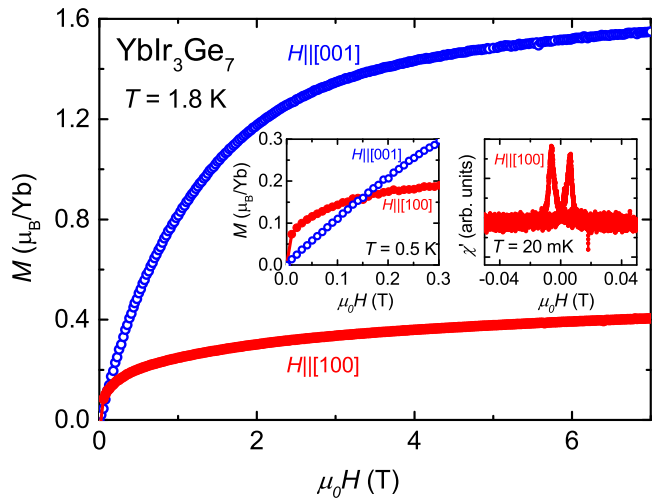}
\caption{\label{MH} Magnetization isotherm $M(H)$ at $T$ = 1.8 K for $H~\parallel$ [100] (closed red circles) and  $H~\parallel$ [001] (open blue circles). Left inset: low field $M(H)$. Right inset: Ferromagnetic hysteresis in the AC susceptibility at 20 mK along $H~\parallel$ [100].}
\end{figure}

A closer look at the ordered state with $T$ = 1.8 K magnetization isotherms (Fig. \ref{MH}) confirms the $H\parallel[100]$ ferromagnetic ordering (red closed circles), while the magnetization shows crossing around $\mu_0H~=0.1$\,T when H$\parallel$[001] (blue, open symbols) and H$\parallel$[100] (red, full symbols): small spontaneous magnetization (left inset) is observed for $H\parallel[100]$, while the $H\parallel[001]$ $M(H)$ is nearly linear at low H. A small magnetization hysteresis with a coercive field $\approx$ 6 mT is revealed at 20 mK in AC susceptibility measurements with $H~\parallel$ [100], best illustrated in the $\chi'(H)$ plot (right inset). These features indicate that the FM ordering occurs with moments along the CEF hard direction [100]. The field along the CEF easy direction [001] rotates the moments to saturation without increasing much above 5\,T, suggesting the absence of a relevant Van Vleck contribution \cite{Lausberg2013}. The saturation magnetization of the ground state Kramers doublet for $H\parallel[100]$ is reached at very small fields with $M^{\mathrm{sat}}_{[100]}\approx0.41\,\mu_{B}$ whereas for $H\parallel[001]$ is reached at fields larger than 4\,T with $M^{\mathrm{sat}}_{[001]} \approx 1.55\mu_{B}$. This yields a relatively large anisotropy factor of about 4. Therefore, assuming anisotropic exchange interaction to explain the FM ordering with moments along the CEF hard axis \cite{Andrade2014} seems unlikely in the case of YbIr$_3$Ge$_7$, because this would necessitate an extremely large anisotropy for the exchange interaction \cite{Hamann2018}.

\begin{figure}[htbp]
	\includegraphics[clip,width=\columnwidth]{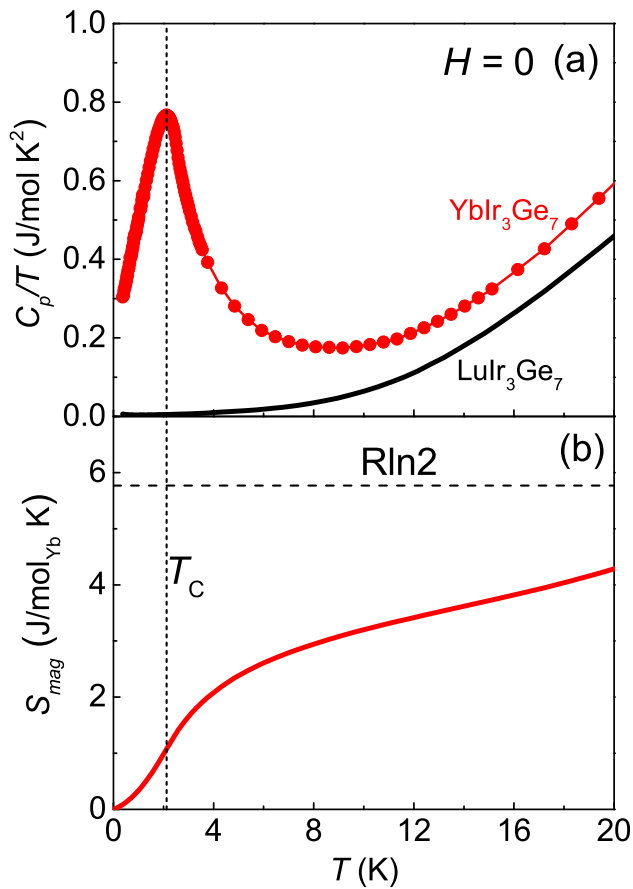}
\caption{\label{Cp} (a) $H$ = 0 specific heat of YbIr$_{3}$Ge$_{7}$ (red symbols and line) and LuIr$_{3}$Ge$_{7}$ (black line) (b) Magnetic entropy of YbIr$_{3}$Ge$_{7}$ with $S_{mag}$ =$\int_{0}^{T} \frac{C_{mag}}{T}~dT$, where $C_{mag} = C_p$(YbIr$_{3}$Ge$_{7}) - C_p$(LuIr$_{3}$Ge$_{7}$).}
\end{figure}

Further evidence of the FM order in YbIr$_{3}$Ge$_{7}$ is shown by the specific heat (Fig. \ref{Cp}), marked by the peak at $T_{\mathrm{C}}~= 2.4$\,K, consistent with the magnetization and resistivity derivatives (Fig. \ref{TC}). As $T\rightarrow 0$, an enhanced electronic specific heat coefficient $\gamma_0~\sim~\frac{C_p}{T}~\approx 300$\,mJ/mol$\cdot$K$^2$ is observed in YbIr$_{3}$Ge$_{7}$ (red), while the corresponding $\gamma_0$ for the non-magnetic analog LuIr$_{3}$Ge$_{7}$ (black line) is, as expected, $<$ 5 mJ/mol K$^2$. The mass renormalization and the small magnetic entropy release at $T_{\rm C}$, $S_{mag}~\sim~17\%$ (Fig.~\ref{Cp}(b)), suggest Kondo lattice formation in YbIr$_{3}$Ge$_{7}$, with a Kondo temperature $T_{\rm K}\sim14$\,K estimated from $S_{mag}(0.5\,T_{\mathrm{K}})=0.5~\mathrm{R}\,\mathrm{ln}2$. This $T_{\rm K}$ estimate is in line with the temperature region where the resistivity exhibits Kondo resonance. YbIr$_{3}$Ge$_{7}$  thus appears to be a rare Yb-based KL ferromagnet with hard axis moment ordering, and the first such compound crystallizing in a rhombohedral lattice.


In Yb-based KLs, the development of FM order away from the CEF easy axis has been revealed in several compounds with different structures, and a wide range of $T_{\rm C}$, from 0.15 K in YbNi$_4$P$_2$ \cite{Krellner2011} to 15 K in CeRuPO \cite{Krellner2008}, while $T_{\rm K}$ ranges from 7 K in CeRuPO \cite{Krellner2008} up to 30 K in YbRhSb \cite{Muro2004}. While YbIr$_{3}$Ge$_{7}$ has a three-dimensional crystal structure, YbNi$_{4}$P$_{2}$ is quasi-one-dimensional. This implies that the dimensionality of magnetic interactions and the relative magnitude of the magnetic and Kondo energy scales, \textit{i.e.}, the position in the Doniach phase diagram \cite{Doniach}, have little to no effect on the FM order along the hard axis in these Kondo ferromagnets. 

Kr\"{u}ger et al. \cite{Kruger2014} proposed a theoretical model to account for the hard axis ferromagnetic order: they suggested that magnetic order along the hard axis can maximize the phase space for spin fluctuations in the easy plane, leading to a minimum in free energy. This is supported by the broadness of the specific heat peak (Fig. \ref{Cp}), hinting at the presence of fluctuations, but is brought into question by the nature of the magnetic anisotropy in YbIr$_3$Ge$_7$: The presence of an easy axis along $c$ rather than an easy plane does not fit well into this picture. In fact, in the case of CeTiGe$_3$ \cite{Fritsch2015}, which has a much stronger easy axis anisotropy, the moments do order along the easy axis. Furthermore, a comparison of YbIr$_3$Ge$_7$  with the other isostructural Yb compounds is called for: YbIr$_3$Si$_7$\cite{StavinohaYbIr3Si7} and YbRh$_3$Si$_7$ \cite{Rai2018YbRh3Si7} are both recently discovered highly anisotropic Kondo lattice systems, with hard axis ferromagnetic correlations in the former, and antiferromagnetic ground state in the latter as suggested by neutron scattering, with a small remanent magnetization of about 0.15\,$\mu_\mathrm{B}/$Yb along the [100] direction. Beyond the KLs showing hard axis ordering, several other strongly correlated FMs \cite{Hafner2018} reveal that magnetic order away from the CEF easy axis is not an exception, but rather a frequent enough occurrence to warrant a thorough theoretical investigation. The "1-3-7" compounds (YbIr$_3$Ge$_7$ with ferromagnetic order and YbIr$_3$Si$_7$ with ferromagnetic correlations \cite{StavinohaYbIr3Si7}, together with the antiferromagnet YbRh$_3$Si$_7$\cite{Rai2018YbRh3Si7}) have the Yb ions in the lowest point symmetry (trigonal) of all these KL compounds. With CEF effects inherently tied to the point symmetry of the magnetic moment, the observation of magnetic order away from the easy axis in different point symmetry cases underlines the difficulty of a generalized theory, which is left to a separate thorough theoretical study.




In conclusion, we report the discovery of a KL compound YbIr$_{3}$Ge$_{7}$ that shows FM ordering at $T_{\rm C}$ = 2.4 K, with the moment lying along the CEF hard direction. With a rhombohedral crystal lattice, this material expands this class of systems to include a new crystal structure. With relatively small $T_{\rm C}$ and $T_{\rm K}$, YbIr$_{3}$Ge$_{7}$ is an ideal candidate to study FM QCP by chemical substitution, and to further develop existing theories to explain FM KL compounds. 

BKR, MS, CLH, and EM acknowledge support from the Gordon and Betty Moore Foundation EPiQS Initiative through grant GBMF 4417. E.M. acknowledges travel support to Max Planck Institute in Dresden, Germany from the Alexander von Humboldt Foundation Fellowship for Experienced Researchers. This research is funded in part by a QuanEmX grant from ICAM and the Gordon and Betty Moore Foundation through Grant GBMF 5305 to Binod K. Rai. We thank the DFG for financial support from project BR 4110/1-1. JYC acknowledges support from NSF: DMR-1700030. 

\bibliography{YbIr3Ge7}

\section{Supplementary Materials}
\beginsupplement

\begin{center}
\renewcommand{\arraystretch}{0.5}
\setlength{\tabcolsep}{12pt}
\begin{table}[htbp]
\caption{\label{SCXRD} Crystallographic parameters of YbIr$_3$Ge$_7$ single crystals at $T$ = 299 K ($R\bar{3}c$)}
\begin{tabular}{c|c}
 \hline
	$a$ (\AA) 									&  7.8062(10)   	    \\
	$c$ (\AA)									& 20.621(5)   	   \\	
	$V$ (\AA$ ^{3}$) 								& 1088.2(4)		   \\
	crystal dimensions (mm$^{3}$) 						& 0.02 x 0.04 x 0.06   \\
	$\theta$ range ($^{\circ}$)	    					& 3.6 - 30.4               \\
	extinction coefficient							& 0.000107(13)	    \\
	absorption coefficient (mm$^{-1}$)					& 94.87		     \\		
	measured reflections							& 7380		    \\
	independent reflections 							& 374 			    \\
	R$_{int}$	 								& 0.046		    \\
	goodness-of-fit on F$^2$	  						& 1.20		   \\
	$R_1(F)$ for ${F^2}_o \textgreater 2\sigma ({F^2}_o)^a$	& 0.017		  \\
	$wR_2({F^2}_o)^b$							& 0.033		 \\ \hline
\end{tabular}
$^{a}R_1 = \sum\mid\mid F_o\mid - \mid F_c\mid \mid / \sum \mid F_o \mid$ 
$^bwR_2 = [\sum[w({F_o}^2 - {F_c}^2)^2]/ \sum[w({F_o}^2)^2]]^{1/2}$ 
\end{table}
\end{center}

\begin{center}
\renewcommand{\arraystretch}{1.2}
\setlength{\tabcolsep}{10pt}
\begin{table*}[h!]
\caption{\label{Atomic} Atomic positions, $U_{eq}$ values, and occupancies for single crystals of YbIr$_3$Ge$_7$}
	\begin{tabular}{ c|c|c|c|c|c}
     \hline

Atom                      & x	          		& y	           	 & z	         		 & $U_{eq}$ (\AA$^2$)$^a$	    	 & Occupancy\\ \hline		
Yb	                    & 0			& 0			 &  0			& 0.00652(14)				 & 1\\
Ir	                    & 0.31885(3)		& 0			 & $\frac{1}{4}$	& 0.00321(11) 			           & 1\\
Ge1	                    & 0.54369(8)	           & 0.68054(8)	 & 0.03056(2)	& 0.0050(2)					 & 0.970(3)\\
Ge2     	         &  0			& 0	                      & $\frac{1}{4}$ 	& 0.0048(4)				           & 0.911(6)\\ \hline
  \end{tabular}

$^a$ $U_{eq}$ is defined as one-third of the trace of the orthogonalized $U_{ij}$ tensor.

\end{table*}

\end{center}

\begin{figure}[h!]
\includegraphics[width=\columnwidth,clip]{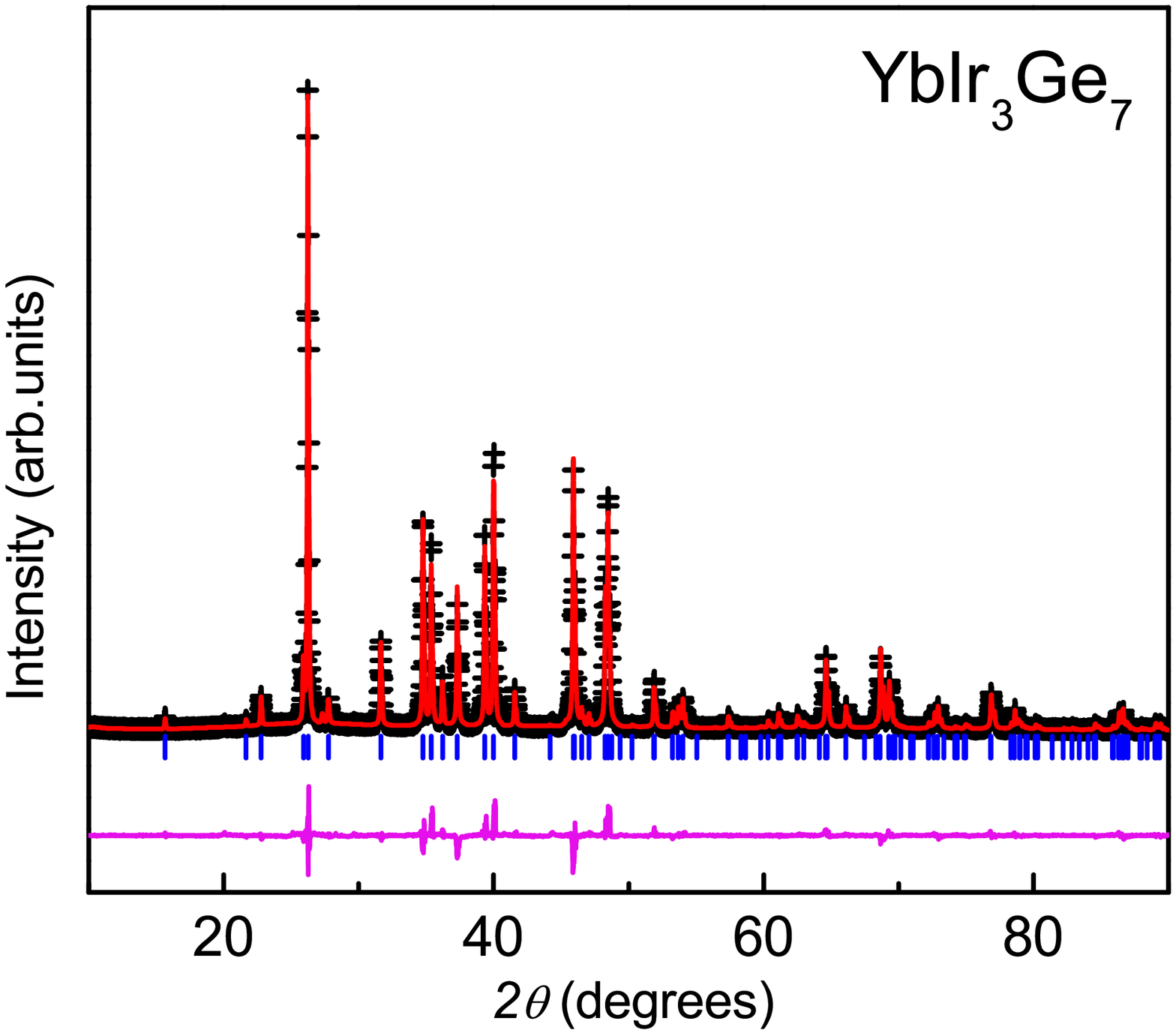}
\caption{\label{XRD} Room temperature powder x-ray diffraction pattern for YbIr$_3$Ir$_7$ (black symbols) together with the calculated pattern (red line), the their difference (violet line) and calculated peak positions (blue vertical lines) using space group $R\bar{3}c$.}
\end{figure}



\end{document}